\documentclass[runningheads]{llncs}
\usepackage[T1]{fontenc}
\usepackage{graphicx}
\usepackage{comment}
\usepackage{url}
\usepackage{amsmath}
\usepackage{color}
\usepackage{makecell}
\begin{document}
\title{Exploring Interactive Simulation of \\ Grass Display Color Characteristic \\ Based on Real-World Conditions}
%
%

\author{Kojiro Tanaka\inst{1,2} \and
Keiichi Sato\inst{1} \and
Masahiko Mikawa\inst{1} \and Makoto Fujisawa\inst{1}}
\authorrunning{K. Tanaka et al.}
\titlerunning{Interactive Simulation of Grass Display Color Characteristic}
%
\institute{ University of Tsukuba, Tsukuba, Ibaraki, Japan \and CyberAgent, Shibuya, Tokyo, Japan \\ \email{tanaka.kojiro.sp@alumni.tsukuba.ac.jp}}

%
%
%
\maketitle              
\vspace{-25pt}
\begin{abstract}
Recent research has focused on incorporating media into living environments via color‑controlled materials and image display. In particular, grass‑based displays have drawn attention as landscape‑friendly interactive interfaces.  To develop the grass display, it is important to obtain the grass color change characteristics that depend on the real environment. However, conventional methods require experiments on actual equipment every time the lighting or viewpoint changes, which is time-consuming and costly. Although research has begun on simulating grass colors, this approach still faces significant issues as it takes many hours for a single measurement. In this paper, we explore an interactive simulation of a grass display color change characteristic based on real-world conditions in a virtual environment.  We evaluated our method’s accuracy by simulating grass color characteristics across multiple viewpoints and environments, and then compared the results against prior work.  The results indicated that our method tended to simulate the grass color characteristics similar to the actual characteristics and showed the potential to do so more quickly and with comparable accuracy to the previous study.

\end{abstract}
\vspace{-30pt}
\section{Introduction}
\vspace{-10pt}
There has been increasing research in integrating media into living environments by controlling colors and displaying images using everyday materials \cite{yamamoto2023turning,ishii2022designing,yu2023thermotion,ishii2019bubbowl,sareen2023bubbletex,nagafuchi2020polka,tsujimoto2016ketsuro,nakanishi2019granular,tokuda2017programmable,robinson2019sustainabot,kimura2014moss,gentile2018plantxel,govonlineRicePaddy}. Among these materials, using grass as a medium has gained attention as it blends naturally into green spaces. Grass has been used in art \cite{ackroydandharveyDearEarth,okayamaartsummit}, sports \cite{mlb}, and advertising \cite{itsnicethatCocaColaTraces}. Recently, a grass display capable of displaying 8-bit level images and animations has been proposed \cite{tanaka2024programmablegrass}. The grass interactive interface is expected to enhance usability while creating a natural green ambiance.

The color characteristics that represent the color changes of media using everyday materials, including the grass display, are essential for evaluating their performance and accurately controlling their colors.  In particular, since material-based media present color by reflecting external light, it is necessary to evaluate not only the physical properties of the material but also the perceptual color as seen by humans, which depends on combinations of viewing positions and illumination environments.  

Currently, these characteristics are mainly measured in real environments \cite{yamamoto2023turning,ishii2022designing,nagafuchi2020polka,tanaka2024programmablegrass}, which require significant time and effort.  This is because each change in viewing position demands camera realignment, and variations in lighting conditions require separate measurements. As the combinations of these conditions increase, the number of required measurements grows rapidly, leading to escalating time, labor, and equipment costs.  Recent research has explored simulating the grass colors using 3DCG (Three-Dimensional Computer Graphics) software with offline rendering and a genetic algorithm \cite{mizuno20233d}. However, the current method, while performing precise evaluations, requires extensive computation time for each measurement.

\vspace{-15pt}
\begin{figure}
  \centering
  \includegraphics[width=\linewidth]{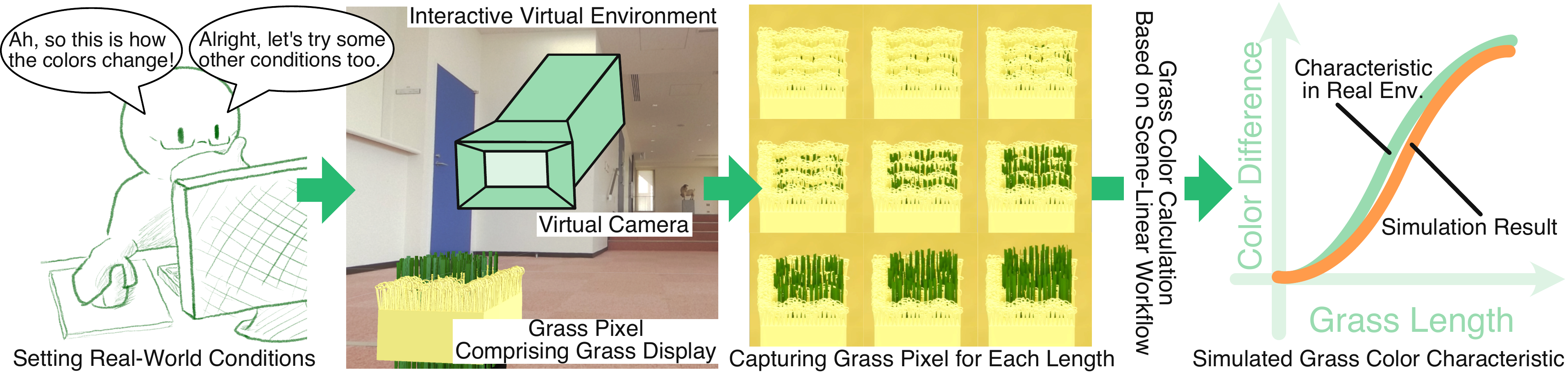}
  \vspace{-20pt}
  \caption{Overview of Process for Interactive Simulation of Grass Color Characteristic Based on Real-World Conditions}
  \label{teaser}
  \vspace{-20pt}
\end{figure}

In this paper, we explore an interactive simulation of a grass display color characteristics as shown in Figure \ref{teaser}. Our method aims to achieve a balance of reasonable accuracy and the ability to interactively modify real-world conditions. We used an interactive virtual environment with simplified light simulation compared to offline rendering. Additionally, we designed and implemented a calculation process capable of simulating real-world grass color characteristics based on a physically-based, scene-linear workflow.

We verified the accuracy of our simulation method through multi-viewpoint, multi-environment comparisons with grass color characteristics measured in real environments. As a result, our proposed method demonstrated the potential to simulate grass color characteristics with reasonable accuracy across multiple viewpoints and environments. Furthermore, we conducted comparative experiments with the previous study that simulated the grass color characteristics using offline rendering and genetic algorithms. The results indicated that our proposed method has the potential to be faster while maintaining accuracy comparable to that of the previous study.

The contributions of this study are summarized as follows:
\begin{itemize}
  \item Design and implementation of an interactive, physically-based simulation for grass-display color-change characteristics. 
  \item Verification of how similar the grass color characteristics simulated by our method are to those in real environments under multiple viewpoints and environments.
  \item Conducting comparative experiments between simulations of our method and the previous study to verify the balance between accuracy and speed.
\end{itemize}
\vspace{-20pt}
\section{Related Work}
\vspace{-10pt}
\subsection{Media of Everyday Materials Including Grass}
 Recent research has focused on media that use the colors of everyday materials to display images. For example, studies have been conducted using carpets \cite{yamamoto2023turning}, bubbles \cite{ishii2019bubbowl,sareen2023bubbletex}, wet grounds \cite{nagafuchi2020polka}, condensation \cite{tsujimoto2016ketsuro}, foodstuffs \cite{robinson2019sustainabot}, wood \cite{smoothwareDanielRozin},  tatami \cite{Kitamura_2025}, electrolysis \cite{ishii2022designing}, particles \cite{nakanishi2019granular}, droplets \cite{tokuda2017programmable}, and thermochromic paint \cite{yu2023thermotion} to display images and animations. The development of these diverse material-based image display methods has advanced media that integrate seamlessly into our daily lives.

In addition, plant materials have been used to develop novel green-based media. Gentile et al. showed the potential of sensitive plants as an image display medium \cite{gentile2018plantxel}. Sensitive plants have the features of closing their leaves when stimulated, and this study adjusted the opening and closing of the leaves using air pressure. Kimura et al. proposed a medium that utilizes moss that opens when given water and closes when dry \cite{kimura2014moss}. In the field of art, there are efforts to draw illustrations in rice paddies using differences in the color of rice plants \cite{govonlineRicePaddy}.

Among media using plants, grass has been particularly noted for its versatility. Some examples of art \cite{okayamaartsummit,ackroydandharveyDearEarth}, sports \cite{mlb} and advertisements \cite{itsnicethatCocaColaTraces} using grass to display illustrations exist. The classical method of creating an image in grass materials is to mow them, however, techniques for grass media have developed, allowing for image display through tip adjustment without mowing grass \cite{newgroundtechnologyAirPrintGently,sugiura2017grassffiti} and realistic expression \cite{ackroydandharveyDearEarth}. Recently, Tanaka et al. proposed a grass display that plays animation \cite{tanaka2021natural,tanaka2023dynamic} and a method of controlling grass colors at 8-bit levels \cite{tanaka2024programmablegrass}. These methods use two types of artificial grass with different colors, one with fixed grass length and the other with adjustable grass length. By adjusting the grass length, it is possible to dynamically control the grass colors.  Our research relates to these grass-based media technologies, specifically focusing on color performance evaluation.

\vspace{-10pt}
\subsection{ Color Performance Evaluation of Material-Based Media}

To display images and animations using everyday materials, including grass, it is important to evaluate how well the colors can be changed. The observed color changes can then be used to check the display contrast and assist in controlling the colors. These evaluation methods are based on various physical properties of the material. For example, there is an evaluation that focuses on a simplified measurement of the material’s color change itself \cite{ishii2022designing}, and another that measures the area over which the color changes on the surface \cite{nagafuchi2020polka}. Among these, there are material-based media whose display performance is evaluated in a perception-aware way by measuring colors based on both lighting environments and the positions of viewpoints. For example, such approach has been applied to evaluating the display performance of carpets \cite{yamamoto2023turning} and grass \cite{tanaka2024programmablegrass}.

However, many evaluations of material-based media are manually performed and labor-intensive. In particular, perception-aware color measurements must take into account not only the properties of the material but also a wide range of viewing positions and a variety of lighting environments.  As a result, the number of measurement conditions increases. For example, even under fixed lighting conditions, prior studies \cite{yamamoto2023turning,tanaka2024programmablegrass} required dozens of measurements across multiple viewpoints and media characteristics, resulting in a significant amount of time and effort to comprehensively evaluate the media’s performance.  The study on carpet-based media \cite{yamamoto2023turning} has pointed out the importance of improving measurement efficiency for color evaluation under multiple conditions. Indeed, there is recent research that evaluates media based on tatami materials by automatically adjusting a single light source and the position of a viewing direction using a precisely calibrated experimental device \cite{Kitamura_2025}. While this evaluation method is designed for use in precisely controlled color evaluation environments, there are few methods for efficiently evaluating perceived colors based on material-based media under diverse lighting conditions and viewing positions in real-world, everyday settings.

Therefore, we focus on the interactive virtual environment created to flexibly reflect real-world conditions and stably obtain the characteristics of the color changes. In this paper, as one of the displays using everyday materials, we aim to simulate the color characteristics of the grass display proposed by Tanaka et al. \cite{tanaka2024programmablegrass}. As an example of the cost of manual measurement of the grass display, measuring each of the 16 viewpoints took approximately five minutes, resulting in a total of about 80 minutes. Our approach aims to reduce the per-condition measurement time to just a few seconds using an interactive virtual environment, while maintaining measurement accuracy, so that the total cost remains low even when multiple conditions are involved. Details of the manual measurement time are provided in Table \ref{tabletime} in Sub-section \ref{multiview}. 

\vspace{-10pt}
\subsection{Simulating Colors in Real Environment with Interactive Virtual Environment}
According to the research of the grass display \cite{tanaka2024programmablegrass}, the color characteristics of the grass display are determined with the grass colors measured by a digital camera. The intensity of light received by the camera is calculated in linear RGB values without gamma correction. This method, called a scene-linear workflow, enables accurate color processing based on the physical properties of light \cite{ibraheem2012understanding,walker2021color,selan2012cinematic}.

In recent years, interactive virtual environments have also started to simulate light using the scene-linear workflow, making it possible to obtain rendered images closer to the real world \cite{unitylinear}. However, color reproduction focusing on the real environment with interactive virtual environments mainly centers on matching the colors seen in the real environment with those seen through devices such as HMDs (Head-Mounted Displays) \cite{diaz2020spectral,diaz2021real,murray2022luminance,rodriguez2024color,patel2024lightness,knecht2011adaptive,menk2012truthful,takezawa2019material,itoh2015semi,zhong2021reproducing,matsuda2022realistic,chen2022gloss,gil2022colour,pardo2018correlation}. Although simulating the color changes of the grass display requires reproducing the linear RGB changes with light physical properties associated with real-world color changes within an interactive virtual environment, there are not many methods for the purpose. 

In related studies, Díaz-Barrancas et al. proposed methods that incorporated the real environment into the virtual environment by spectrally analyzing light sources and object surfaces \cite{diaz2020spectral,diaz2021real}. However, since these methods require spectral analysis of light sources, they are limited to simple lighting conditions, in addition to the high cost of spectral analysis devices. This makes them unsuitable for simulating the color characteristics of the grass display in multiple real environments.

Although it is not in the interactive environment, Mizuno et al. proposed a method to simulate the colors of the grass display using a 3DCG software for offline rendering with a linear workflow and a genetic algorithm \cite{mizuno20233d}. This method uses an HDRI (High Dynamic Range Image) as the lighting condition and employs a genetic algorithm to determine the albedo color of the grass itself, using the real images of the grass materials as the ground truth. This method showed the potential for the grass colors obtained in the CG space to approximate the grass colors measured in the real environment. However, this method requires many hours for measurements every time the viewing position or lighting environment is changed due to the influence of the offline rendering and the genetic algorithm. 

In this study, we design and implement a simulation of color characteristics of a grass display using an interactive virtual environment that allows for simpler light simulation compared to offline rendering, based on a scene-linear workflow. Our method is developed, focusing on a balance of reasonable accuracy and the ability to interactively modify real-world conditions, avoiding high-precision techniques such as spectral analysis and genetic algorithms. We verify how well our method, which prioritizes interactive simulation over accuracy, can simulate the grass display color characteristics based on the real environment through multi-viewpoint, multi-environment, including outdoor environments, and a comparative experiment with the conventional method.

\vspace{-10pt}

\section{Design and Implementation}
\vspace{-10pt}
 We aim to simulate how the color changes of the grass display are perceived under multiple real-world lighting and viewing conditions.  In this section, we explain the design and implementation of the virtual environment used to simulate the color characteristics of the grass display. We used Unity HDRP (High Definition Render Pipeline), which enables physically-based real-time rendering, as the basis for constructing the virtual environment \cite{unity}.
\vspace{-10pt}
\subsection{Structure of Grass Display}
\begin{figure}[h]
  \centering
  \includegraphics[width=0.8\linewidth]{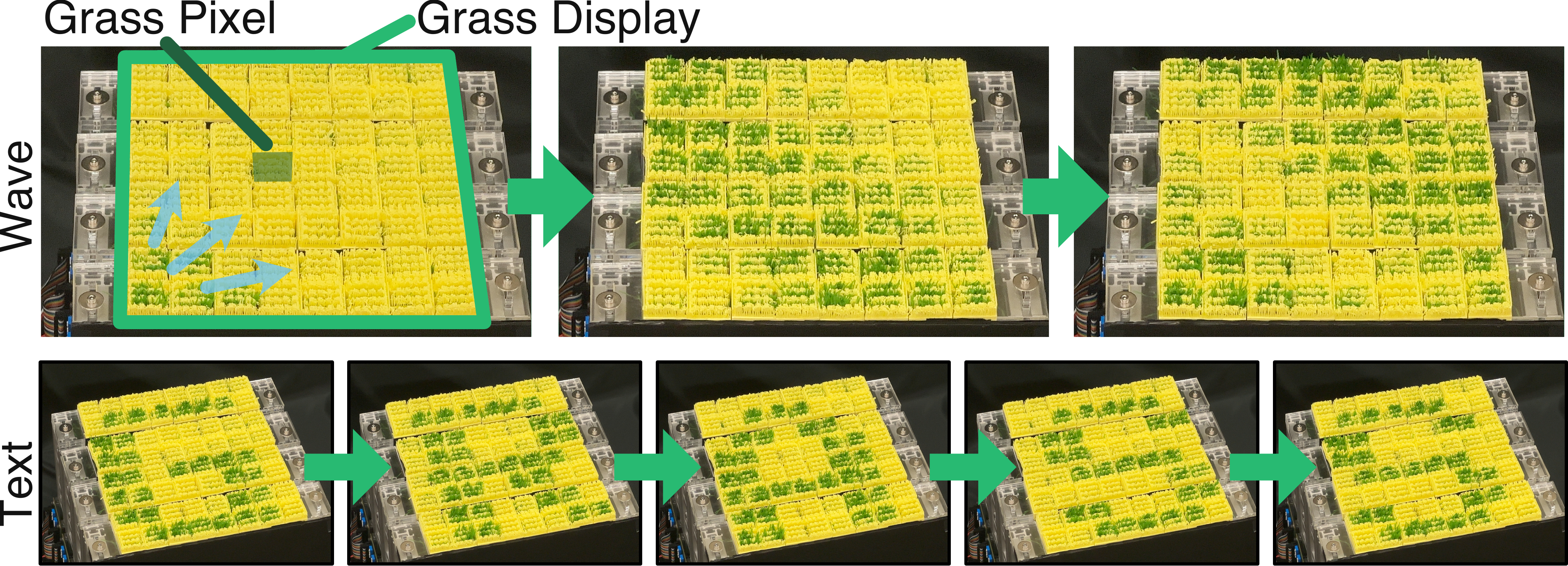}
\vspace{-10pt}
  \caption{Grass Display Playing Wave and Text Animations} 
  \label{grassdisplay}
\end{figure}

The grass display controls the grass color by combining two types of grass: one with fixed length and the other with adjustable length on a pixel-by-pixel basis; these pixels are called grass pixels. Figure \ref{grassdisplay} shows an $8\times8$ grass display constructed with fixed-length yellow artificial grass and adjustable-length green artificial grass. As the length of the grass increases, the area occupied by the green color gradually increases compared to the yellow. At this time, due to the fine texture of the grass, the human retina can hardly distinguish individual grass colors, so humans perceive the overall color of the grass as an average color, a phenomenon known as spatial additive color mixing \cite{tanaka2024programmablegrass}.

The color characteristics of the grass display are represented by an origin grass color difference (OGCD). An OGCD is a color difference between the grass color when the adjustable grass length is at its minimum and the color at a given grass length. Thus, the relationship between the OGCDs and the grass lengths expresses the color characteristics of the grass display, and in this paper, is defined as grass color characteristics. The calculation method of the OGCD is explained in detail in Subsection \ref{subsec_colorprocess}. The grass color characteristics are influenced not only by the grass display components but also by the lighting environments and the viewpoint positions.

\vspace{-10pt}
\vspace{0pt}

\vspace{-30pt}

\subsection{Materials Required for Virtual Environment}

This subsection describes the materials required to simulate the grass color characteristics in the virtual environment. Figure \ref{IO} illustrates the materials needed primarily for replicating the lighting conditions of the real environment and for constructing 3D models of a grass pixel and a color checker. In this study, a grass pixel is used to evaluate the grass color characteristics of the grass display.

\vspace{-10pt}

\subsubsection{Lighting Configuration}

\begin{figure}[h]
\vspace{-15pt}
  \centering
  \includegraphics[width=0.6\linewidth]{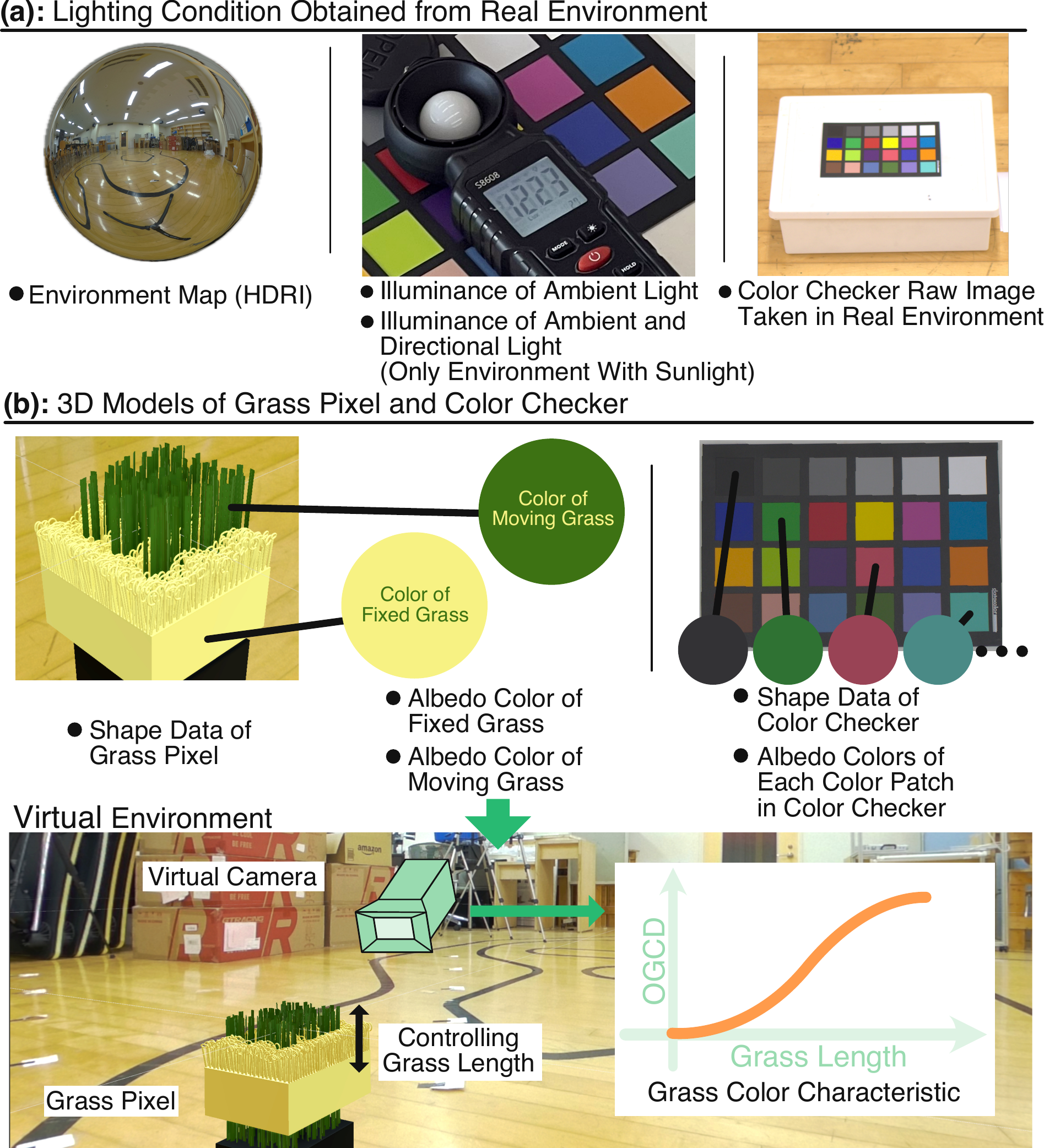}
\vspace{-10pt}
  \caption{Configuration of Virtual Environment} 
  \label{IO}
  \vspace{-40pt}
\end{figure}

Figure \ref{IO}(a) shows the materials required to replicate the lighting conditions of the real environment. An ambient light environment map of an HDRI \cite{debevec2010high} captured with an omnidirectional camera is used. The Unity HDRP allows defining the brightness of HDRI as a physically-based light unit. The illuminance [Lux] at the position where the omnidirectional camera is used is measured, and the HDRI intensity is adjusted accordingly.

 Since accurately reproducing direct sunlight in real‑time rendering using HDRI alone is challenging, we supplement it with a directional light source.  Thus, a directional light object is added in addition to HDRI,  with ray-traced shadows enabled for direct lighting.  In addition, the illuminance at the position of the omnidirectional camera is measured both when the sunlight is blocked and when it is not blocked. The illuminance measured without sunlight is considered the illuminance of the ambient light, which is used to adjust the HDRI intensity. The difference between the illuminance with and without sunlight is considered the sunlight illuminance, and the intensity of the directional light object is adjusted accordingly. Additionally, to determine the direction of the directional light object, a metal ball is placed in the virtual environment that reflects the environment map of the HDRI, allowing the direction of the sunlight to be seen and the directional light can be adjusted accordingly.

As a color reference for the real environment, a color checker is placed at the position where the omnidirectional camera is used and photographed as a raw image with a digital camera with adjusted exposure. This color checker is necessary to align the linear RGB values of the virtual environment with those of the real environment. The color calculation process using this color checker is explained in Subsection \ref{subsec_colorprocess}. Thus, the lighting conditions of the real environment are reflected in the virtual environment.

\subsubsection{3D Model Configuration}

\begin{figure}[h]
  \vspace{-10pt}
  \centering
  \includegraphics[width=0.6\linewidth]{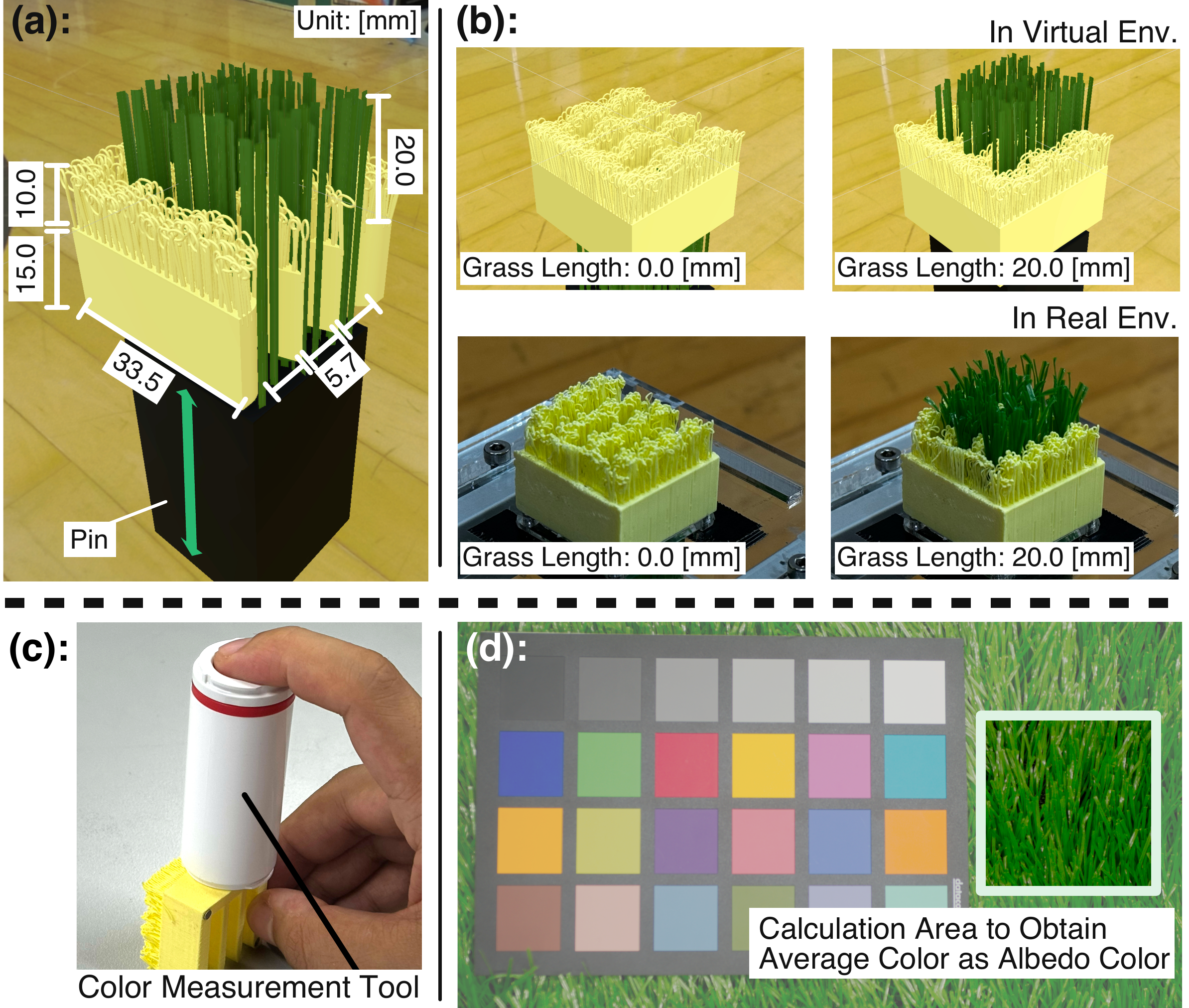}
  \vspace{-10pt}
  \caption{Grass Pixel in Virtual and Real Environments, and Measuring sRGB of Grass} 
  \label{grasspixel}
    \vspace{-20pt}
\end{figure}

Figure \ref{IO}(b) illustrates the 3D models required for the virtual environment, including the grass pixel and the color checker. These 3D models are created based on their actual shapes. Figure \ref{grasspixel}(a) shows the grass pixel shape used in this paper. Yellow grass represents fixed-length grass, and green grass represents adjustable-length grass. The grass pixel's surface area is $33.5\times33.5$ [mm], with a base height of 15.0 [mm]. Fixed-length grass of 10.0 [mm] is planted on the base surface, which has three slits, each 5.7 [mm] wide, for adjustable-length grass. The adjustable-length grass is planted on a pin beneath the fixed-length grass, allowing the length to vary from 0.0 to 20.0 [mm] relative to the base height. Figure \ref{grasspixel}(b) shows the comparison between the grass pixels in the virtual and real environments. In this study, the real grass pixel consisted of fixed-length grass made by 3D printing and adjustable-length grass made from commercial artificial grass. In addition, the density of grass in the 3D-modeled grass pixel is adjusted to align with the density of the real fixed-length and adjustable-length grass.

Next, the method for determining the colors of the 3D model is explained. In the Unity HDRP, the color of an object is primarily determined by its albedo color, metallic, and smoothness properties \cite{unitymaterial}. Since the color of the grass pixel is determined through spatial additive color mixing with two different original grass colors, the albedo color is the factor that most significantly influences the grass color characteristics. Thus, this study focuses on the albedo color to simulate the basic color characteristics of the grass pixel while using the default values for metallic and smoothness in the Unity HDRP, setting the metallic parameter to zero and the smoothness parameter to a mid-level value.

The albedo color represents the reflectance of the object. To accurately obtain the reflectance of the object in the real environment, a hyperspectral camera is typically required \cite{diaz2020spectral,diaz2021real}, which can be expensive. As a simpler alternative, this study uses a color measurement tool based on a standard color evaluation environment. This tool can measure the color of a flat surface in about several seconds and provides the color in an sRGB value, which is used as the albedo color for the object. The Datacolor ColorReader CR100 is used as the color measurement tool in this study.

The albedo colors of the fixed-length grass and the color patches on the color checker are determined using this color measurement tool on the flat surfaces of the actual objects as shown in Figure \ref{grasspixel}(c). However, since the adjustable-length grass consists of fine blades and lacks a flat surface, it is difficult to measure its color with the color measurement tool. Thus, its albedo color is determined in a standard color evaluation environment as ISO 3664:2009 \cite{iso}, photographing an image adjusted for exposure and white balance using the color checker, and calculating the average color of a section of the grass in a sRGB value as shown in Figure \ref{grasspixel}(d). Therefore, these procedures allow the grass pixel and the color checker to be modeled in the virtual environment.

\subsection{Calculation of Grass Color Characteristic}
\label{subsec_colorprocess}

\begin{figure*}[h]
\vspace{0pt}
\centering
\includegraphics[width=\linewidth]{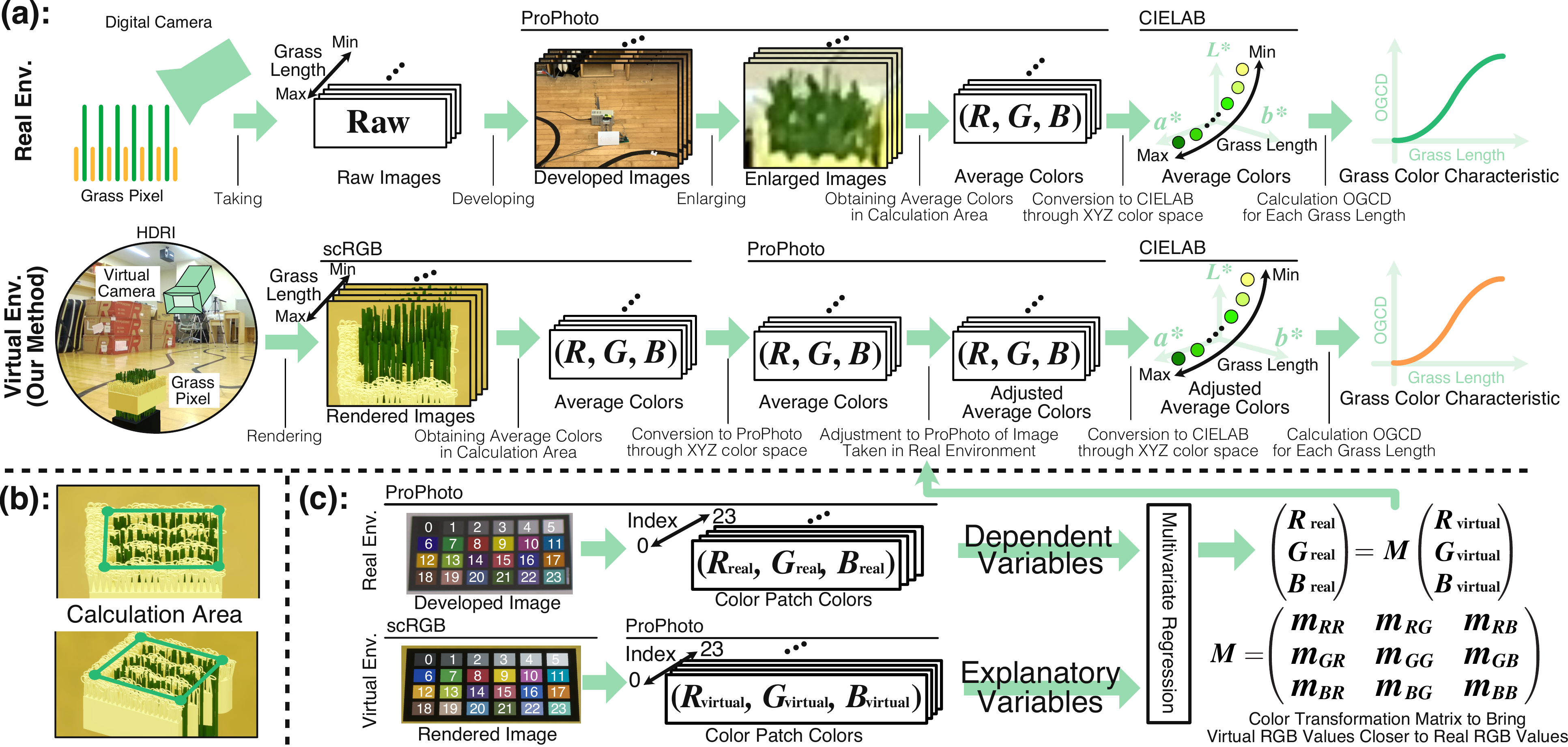}
  \vspace{-20pt}
\caption{Calculation for Grass Color Characteristics in Real and Virtual Environments}
\label{colorprocess}
\vspace{-15pt}
\end{figure*}

This subsection explains the calculation process of the grass color characteristic in the real and virtual environment, as shown in Figure \ref{colorprocess}(a).

\subsubsection{Process in Real Environment}
First, we describe the calculation method in the real environment, as proposed by Tanaka et al \cite{tanaka2024programmablegrass}. A grass pixel and a digital camera with exposure adjusted using an 18\% gray card are prepared. The grass length is varied from minimum to maximum at regular intervals, and a raw image is captured each time. The raw images are developed into uncompressed tiff images in linear RGB color space within a wide gamut of ProPhoto \cite{prophoto}, with white balance adjustments. The developed images are enlarged to focus on the grass pixel. The calculation area for the grass color is then defined, as shown in Figure \ref{colorprocess}(b), where the area is enclosed by lines connecting the four corners of the grass pixel. The average color within this area, specifically the average values of the R, G, and B channels, is calculated.

To derive the grass color characteristic, the average linear RGB values are converted through the CIE (Commission internationale de l'éclairage) XYZ color space to obtain the CIELAB (CIE $L^*a^*b^*$) values for each grass length. By using the CIEDE2000 color difference formula $\Delta E^*_{00}$, the color difference between two CIELAB values \((L_1^*, a_1^*, b_1^*)\) and \((L_2^*, a_2^*, b_2^*)\) can be obtained in a way that aligns with human perception \cite{sharma2005ciede2000} as below:

\[
\Delta E^*_{00}((L_1^*, a_1^*, b_1^*), (L_2^*, a_2^*, b_2^*))
\]

\noindent Let \((L^*, a^*, b^*)_{\text{Length}=x}\) represent the CIELAB value at a certain grass length \(x\). To obtain the grass color characteristic, the CIELAB value at the minimum grass length is used as a reference to calculate the color difference for each grass length, defined as OGCD (Origin Grass Color Difference) as below. The grass color characteristic can be determined by calculating the OGCD for each grass length and fitting a curve to the relationship between the OGCD and the grass length.

\vspace{-10pt}

\[
\text{OGCD}(x) = \Delta E^*_{00} \left( (L^*, a^*, b^*)_{\text{Length=min}}, (L^*, a^*, b^*)_{\text{Length}=x} \right)
\]

\vspace{-10pt}

\subsubsection{Process in Virtual Environment}

Next, we explain the calculation of the grass color characteristic in the virtual environment including the grass pixel and a virtual camera. The grass length varied from minimum to maximum, and the scene is captured using the virtual camera to render images. These rendered images are obtained in scRGB format \cite{scRGB}, which is a linear RGB color space in the Unity HDRP. Similar to the real environment, the calculation area for the grass pixel is defined as shown in Figure \ref{colorprocess}(b), and the average scRGB values are obtained from this area.

The acquired scRGB values are converted through the CIE XYZ color space to the ProPhoto linear RGB values. Subsequently, the converted linear RGB values are adjusted to match the measured linear RGB values of the real environment. This RGB adjustment is performed using a $3\times3$ color transformation matrix \(\textbf{\textit{M}}\), the derivation of which is shown in Figure \ref{colorprocess}(c). The color checker with 24 color patches is used. In the real environment, the color checker is captured, and the average color of each color patch is determined in the ProPhoto linear RGB value. In the virtual environment, the 3D-modeled color checker is rendered, and the average color of each color patch is obtained in the scRGB value and converted to the ProPhoto linear RGB value. Then, multivariate regression is performed, with the real environment's color patches as the dependent variables and the virtual environment's color patches as the explanatory variables, to derive the color transformation matrix \(\textbf{\textit{M}}\). The relationship between the real and virtual environment colors through the matrix \(\textbf{\textit{M}}\) is given by:

\vspace{-10pt}

\[
\begin{pmatrix}
R_{\text{real}} \\
G_{\text{real}} \\
B_{\text{real}}
\end{pmatrix}
=
\textbf{\textit{M}}
\begin{pmatrix}
R_{\text{virtual}} \\
G_{\text{virtual}} \\
B_{\text{virtual}}
\end{pmatrix}
\quad \text{where} \quad
\textbf{\textit{M}} =
\begin{pmatrix}
m_{RR} & m_{RG} & m_{RB} \\
m_{GR} & m_{GG} & m_{GB} \\
m_{BR} & m_{BG} & m_{BB}
\end{pmatrix}
\]

\noindent Here, \((R_{\text{real}}, G_{\text{real}}, B_{\text{real}})\) and \((R_{\text{virtual}}, G_{\text{virtual}}, B_{\text{virtual}})\) represent the ProPhoto linear RGB values in the real and virtual environments, respectively, obtained from the color patches. The elements \(m_{RR}\), \(m_{RG}\), \(m_{RB}\), etc., of the color transformation matrix \(\textbf{\textit{M}}\) represent the coefficients that map the virtual environment's RGB values to the real environment's RGB values. 

Using this matrix, the linear RGB values in the virtual environment can be matched to those in the real environment. Finally, similar to the process in the real environment, the obtained linear RGB values are converted to the CIELAB values, and the OGCDs are calculated to derive the grass color characteristic.

\section{Experiments of Grass Color Characteristic Simulation}

To verify how accurately the grass color characteristics calculated by our virtual environment simulate those in the real environment, several experiments were conducted. In the real environment, an iPad Pro (f/1.8, wide-angle lens, resolution: $4024\times3016$ [pixel]) was used as the digital camera, and in the real grass pixel, the colors of the fixed-length and adjustable-length grass were yellow and green, respectively. The grass length was controlled using a DC motor with a rotary encoder, and a PD (Proportional-Differential) controller implemented by a microcontroller (Teensy 4.1). The microcontroller communicated with a Bluetooth module (ESP32) via SPI (Serial Peripheral Interface), enabling wireless adjustment of the grass length from the iPad Pro, allowing for automatic raw image capture while controlling the grass length.

The HDRI was captured using a RICOH THETA Z1 omnidirectional camera. The color checker used was the Datacolor Spydercheckr 24. The computer used to run the virtual environment had the following specifications: CPU: Intel Core i7-13700, GPU: GeForce RTX 4070, RAM: 32 GB. The virtual camera was zoomed in to ensure that the entire grass pixel was visible, and the rendered image resolution was set to $600\times400$ [pixel].

\vspace{-13pt}

\subsection{Experiments of Multiple Viewpoints}
\label{multiview}
\subsubsection{Experimental Setup}

\begin{figure}[h]
  \centering
  \vspace{-15pt}
    \includegraphics[width=0.5\linewidth]{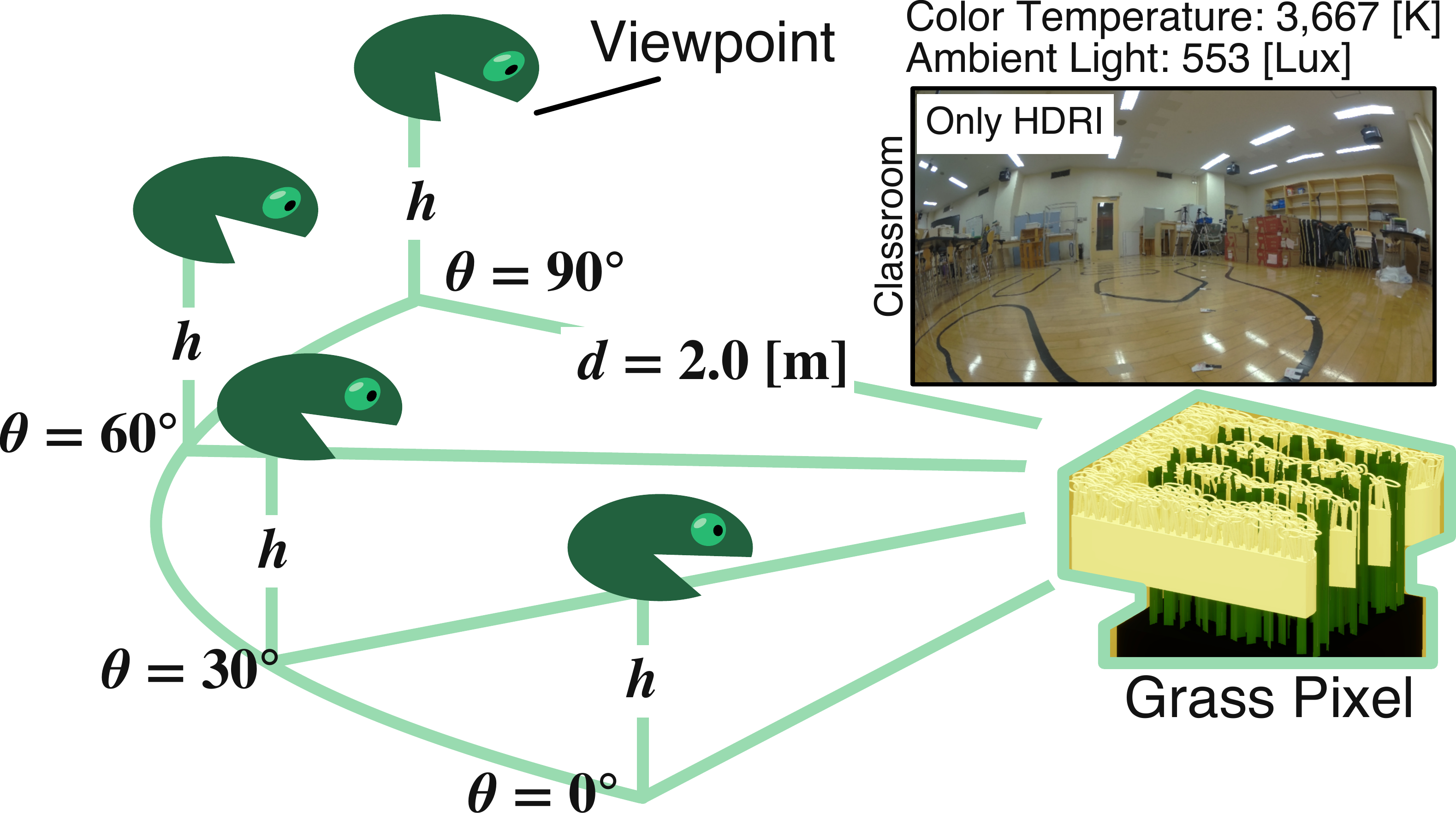}
    \vspace{-10pt}
  \caption{Viewpoint's Positions and Real Environment} 
  \label{position}
  \vspace{-10pt}
\end{figure}

The objective of these experiments is to determine how closely the grass color characteristics from the virtual environment resemble those in the real environment from multiple viewpoints. As shown in Figure \ref{position}, the experiments were conducted in a university classroom with stable lighting conditions.  The viewpoint was determined by the viewpoint's height \( h \), the distance \( d \) between the grass pixel and the viewpoint, and the horizontal angle \( \theta \) relative to the grass pixel. Here, \( \theta = 0^\circ \) was defined as the angle at which the slits of the grass pixel were perpendicular to the viewpoint's line of sight. In these experiments, the distance \( d \) was fixed at 2 [m], and the heights \( h \) of the viewpoints were set to 150, 160, 170, and 180 [cm] to represent the heights of adult males and females. The angles \( \theta \) were set to 0\(^\circ\), 30\(^\circ\), 60\(^\circ\), and 90\(^\circ\), resulting in a total of 16 viewpoints. The position $(h, d, \theta)$ = (170, 2, 0) was used as the reference point to align the colors from the virtual environment to the real environment, and the color correction matrix \(\textbf{\textit{M}}\) derived at this position was then applied to the rendered images viewed from 16 different viewpoints.

We tested the repeatability of the real grass pixel's ability to display its grass colors. The viewpoint was set at $(h, d, \theta) = (170, 2, 0)$, and the trial of the repeatability test was conducted 10 times. In each trial, the grass length was changed from 0.0 to 20.0 [mm] at approximately 1.0 [mm] intervals, and the OGCD was obtained for each grass length. After conducting the 10 trials, the standard deviation of the OGCD for each grass length was calculated, and the average value of these standard deviations was determined. As a result, the repeatability capability of the real grass pixel was found to have an average OGCD standard deviation of approximately 0.42.

\vspace{-10pt}

\subsubsection{Results}
\begin{figure}[h]
  \centering
  \vspace{-10pt}
    \includegraphics[width=\linewidth]{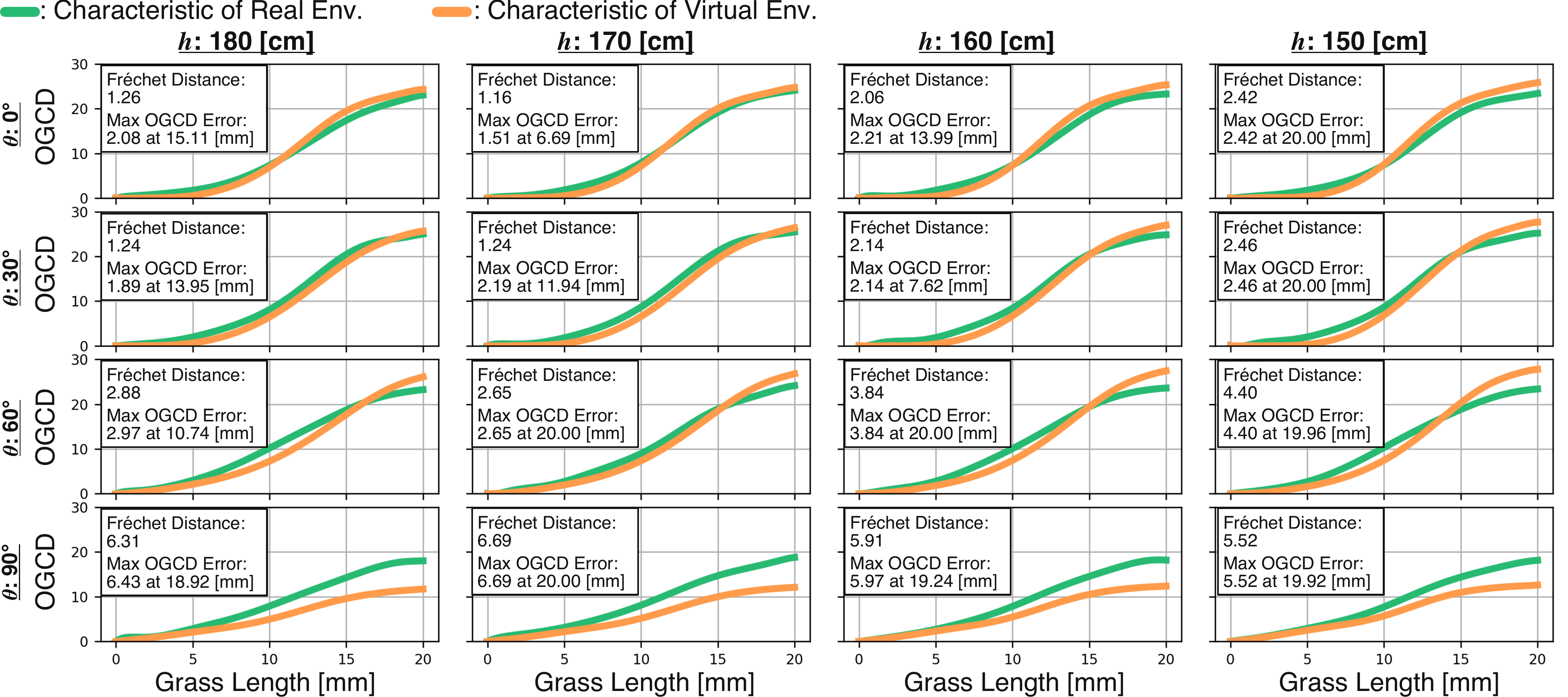}
  \vspace{-20pt}
  \caption{Grass Color Characteristic Results in Real and Virtual Environments for Multiple Viewpoint Positions} 
  \label{charactresult}
  \vspace{-5pt}
\end{figure}

\begin{figure}[h]
  \centering
  \vspace{-10pt}
    \includegraphics[width=0.75\linewidth]{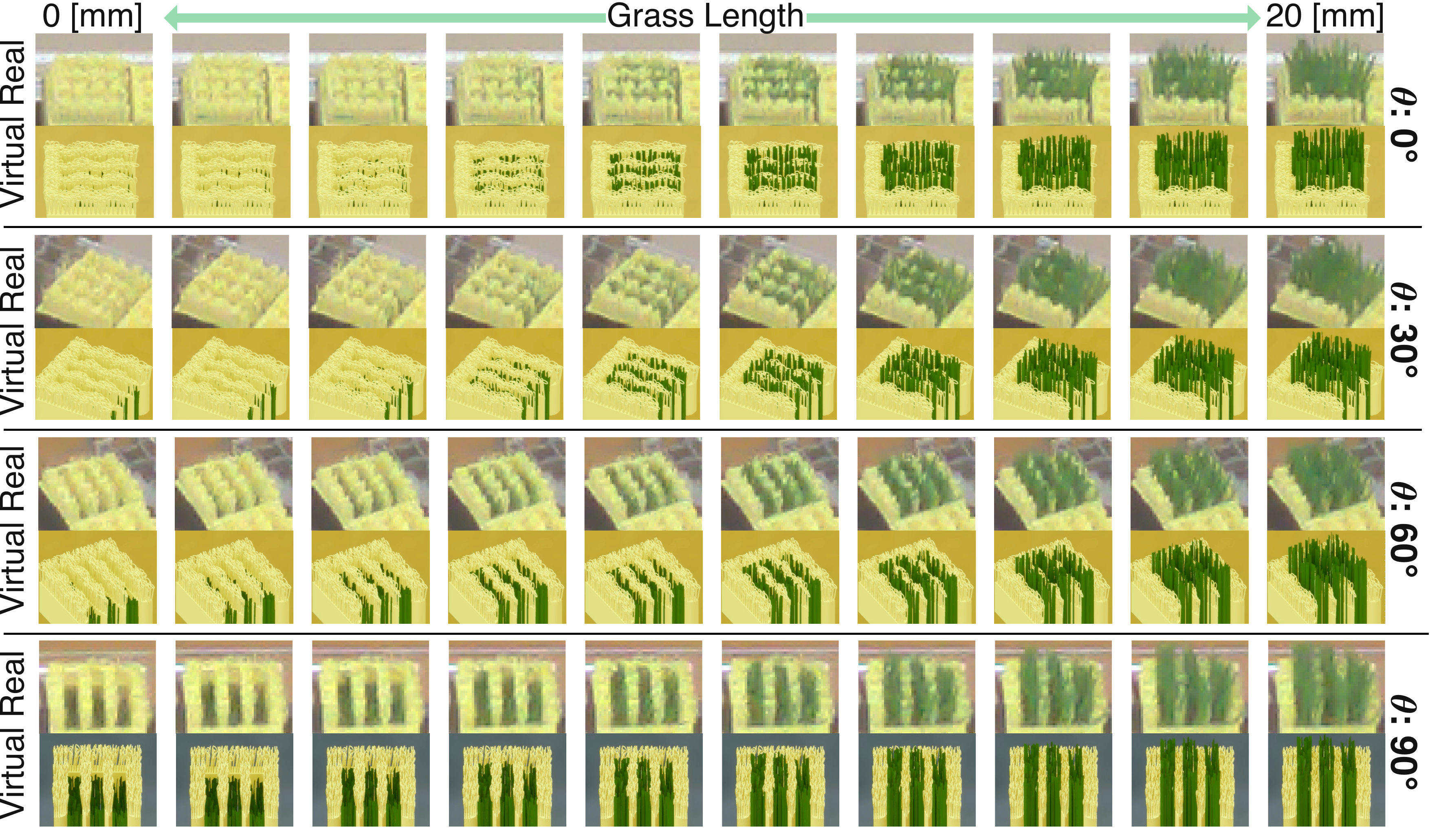}
    \vspace{-10pt}
  \caption{Real and Virtual Grass Pixel Appearance} 
  \label{visualgrass}
  \vspace{-20pt}
\end{figure}

Figure \ref{charactresult} shows the plots of the grass color characteristics from the experiments. In each plot, the vertical axis represents OGCD, and the horizontal axis represents the grass length [mm]. The grass color characteristics obtained from the real and virtual environments are depicted by green and orange curves, respectively. For each plot, a Fréchet distance and a maximum OGCD error were calculated to quantitatively evaluate the similarity between the two curves. Additionally,  Table \ref{tabletime} shows a comparison of the approximate measurement times in these experiments between the manual measurement and our method. Moreover,  the examples of the grass pixel color changes in the real and virtual environments are shown visually in Figure \ref{visualgrass} based on \( h = 170 \) [cm].

\begin{table}[h]
\vspace{-30pt}
\centering
\caption{ Measurement Times Between Manual and  Our Methods}
\label{tabletime}

\begin{tabular}{|c|c|c|}
\hline
 & \textbf{Manual Method} & \textbf{Our Method} \\
\hline
\makecell[c]{Measurement Time \\ per Viewpoint} & \makecell[c]{ About 5 Minutes \\ (Including Camera Adjustment)} & About 5 Seconds \\
\hline
\makecell[c]{Total Time \\ for 16 Viewpoints} & About 80 Minutes & About 80 Seconds \\
\hline
\end{tabular}
\vspace{-10pt}
\end{table}

The experimental results revealed that the grass color characteristics from the virtual environment exhibited similar trends to those measured in the real environment. However, the simulation accuracy of the characteristics varied depending on \( h \) and \( \theta \).

First, focusing on the influence of \( h \), the plots show that the curve of the grass color characteristics simulated at \( h = 170 \) [cm] was the closest to the real environment. For example, when \( \theta \) was \( 0^\circ \), the OGCD ranges were almost the same in the real and virtual environments, with the Fréchet distance of 1.16 and the maximum OGCD error of 1.51 at the grass length of 6.69 [mm]. These values were the smallest among the 16 viewpoints. In contrast, at \( h = 150 \) [cm], the OGCD error was 2.42, the largest OGCD error between the real and virtual environments, with the Fréchet distance of 2.42. The simulation accuracy at \( h = 160 \) [cm] and \( h = 180 \) [cm] was intermediate, with the Fréchet distances of 1.26 and 2.06, and the maximum OGCD errors of 2.08 at the grass length of 15.11 [mm] and 2.21 at the grass length of 13.99 [mm], respectively. This trend was similar at \( \theta = 30^\circ \) and \( \theta = 60^\circ \). However, at \( \theta = 90^\circ \), the Fréchet distance at \( h = 150 \) [cm] was smaller than that at \( h = 170 \) [cm].

Next, the influence of \( \theta \) is considered. In both the real and virtual environments, as \( \theta \) increased, the S-shaped grass color characteristics tended to become a smooth linear shape. This is consistent with the experimental results of Tanaka et al \cite{tanaka2024programmablegrass}. The influence of \( \theta \) showed the same tendency at all \( h \). For example, when \( \theta = 0^\circ \), the simulation accuracy was the highest among all angles \( \theta \), and the OGCD ranges were almost the same in the real and virtual environments. However, when \( \theta \) increased to \( 30^\circ \) and \( 60^\circ \), the OGCD range in the real environment did not change much, while the OGCD range in the virtual environment gradually increased. Furthermore, at \( \theta = 90^\circ \), the simulation accuracy of the characteristics differed greatly compared to other \( \theta \), and the OGCD range in the virtual environment was lower than that in the real environment at all \( h \).

\vspace{-20pt}
\subsubsection{Discussion}
The experiments demonstrated the potential for simulating grass color characteristics based on the real environment under specific conditions. When \( \theta = 0^\circ \), the highest simulation accuracy was observed at \( h = 170 \) [cm]. In contrast, the lowest accuracy was at \( h = 150 \) [cm], with intermediate accuracy at \( h = 160 \) [cm] and \( h = 180 \) [cm]. These results are likely because the color correction matrix \(\textbf{\textit{M}}\) was derived at \( h = 170 \) [cm] and \( \theta = 0^\circ \), making it the reference point for adjusting the linear RGB values of the virtual environment to match those of the real environment.

However, considering that in graphical industry, a CIEDE2000 value around 2.0 is regarded as indicating nearly identical colors \cite{cgats}, the differences in the grass color characteristics between the real and virtual environments may not be significant at all \( h \) values for \( \theta = 0^\circ \) and \( \theta = 30^\circ \), where the maximum OGCD error ranges from 1.16 to 2.46. In contrast, at \( \theta = 60^\circ \) and \( \theta = 90^\circ \), the maximum OGCD error ranges from 2.65 to 6.69, making this less applicable.

As \( \theta \) increased, the OGCD range in the virtual environment tended to rise, contrary to the grass color characteristics in the real environment. However, at \( \theta = 90^\circ \), the OGCD range in the virtual environment significantly decreased compared to the real environment. This is likely due to differences in the physical shape of the green grass. The green grass in the virtual environment extends straight, while in the real environment, the green grass spreads out slightly as it grows. As shown in Figure \ref{visualgrass}, this difference in the grass shapes is less apparent at \( \theta = 0^\circ, 30^\circ \) and \( 60^\circ \). However, at \( \theta = 90^\circ \), the green grass tends to cover the yellow grass more in the real environment compared to the virtual environment, resulting in a stronger green color. This suggests that improving the accuracy of the grass color simulation may require more detailed physical simulations of object shapes, including collision and gravity settings.

\vspace{-10pt}

\subsection{Experiments in Multiple Environments}
\label{submultienv}
\vspace{-7pt}
\subsubsection{Experimental Setup}
We evaluate how accurately the virtual environment can simulate grass color characteristics across multiple environments. The grass color characteristics were measured in both real and virtual environments across five different settings: standard color evaluation environment (Env 1), indoor environment with natural light (Env 2), tree-shaded area (Env 3), brick-paved area (Env 4), and grass field (Env 5).

Env 1 was set up based on ISO 3664:2009 \cite{iso}, which defines color evaluation environments, using fluorescent lamps (Color Temperature: 5000 [K], Color Rendering Index (Ra): 97). The height of the lamps was adjusted so that the illuminance at the grass pixel surface was 2000 [Lux]. In Env $1\sim3$, lighting conditions were constructed using only HDRI. In Env 4 and Env 5, direct sunlight was included, so a directional light object was added in addition to HDRI. The viewpoint position was set at $(h, d, \theta)$ = (170, 2, 0).

\vspace{-15pt}

\subsubsection{Results}

\begin{figure}[h]
  \centering
  \vspace{-10pt}
    \includegraphics[width=\linewidth]{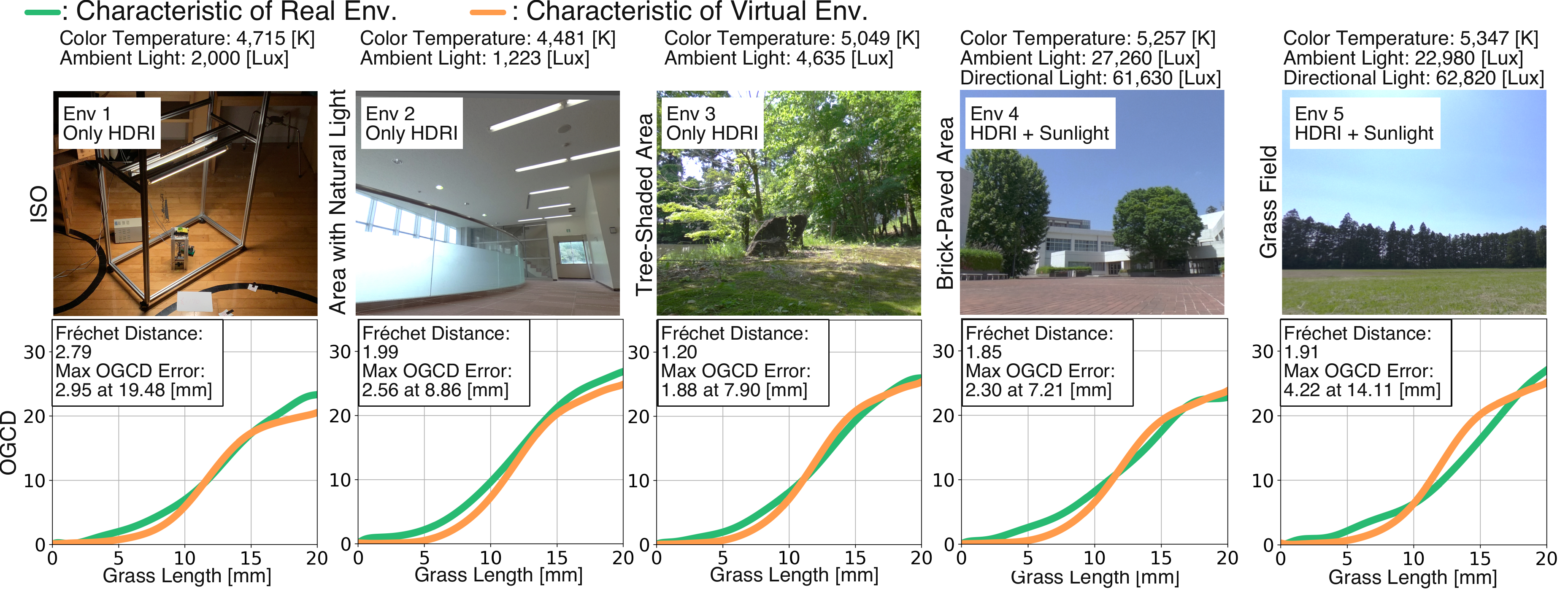}
      \vspace{-20pt}
  \caption{Grass Color Characteristic Results in Multiple Environments}
  \label{multienv}
  \vspace{-20pt}
\end{figure}

Figure \ref{multienv} compares the grass color characteristics in five different environments from the real and virtual environments. The grass color characteristics from the real and virtual environments are represented by green and orange curves, respectively. 

In Env 1, the Fréchet distance was 2.79, and the maximum OGCD error was 2.95 at a grass length of 19.48 [mm]. Generally, the grass color characteristics from the virtual environment showed a similar trend to those of the real environment. However, as the grass length exceeded 15.00 [mm], the increase in OGCD in the virtual environment was more gradual compared to the real environment.

In Env 2 and Env 3, the grass color characteristics from the real and virtual environments were more similar compared to the other environments, with the Fréchet distances of 1.99 and 1.20, and the maximum OGCD errors of 2.56 at the grass length of 8.86 [mm] and 1.88 at the grass length of 7.90 [mm], respectively.

In Env 4, the Fréchet distance was  1.85, and the maximum OGCD error was  2.30  at a grass length of  7.21  [mm]. The grass color characteristics were generally consistent between the real and virtual environments, however, the OGCD values in the virtual environment were smaller than those in the real environment for grass lengths from 0.00 to 7.00 [mm].

In Env 5, the Fréchet distance was  1.91, and the maximum OGCD error was  4.22  at a grass length of  14.11  [mm]. Among the environments tested, this result had the lowest simulation accuracy of grass color characteristics. Particularly, for the grass lengths from 10.00 [mm] to 20.00 [mm], the grass color characteristics in the virtual environment, similar to the other environments, showed a more gradual increase from around 15.00 [mm], whereas in the real environment, the increase continued without becoming gradual.

\vspace{-12pt}

\subsubsection{Discussion}
The HDRI-only environments showed higher simulation accuracy than those using HDRI with a directional light object. Specifically, in Env 2 and Env 3, the OGCD range and characteristics closely matched the real environment. Env 1 also showed a similar curve shape, but the grass color changes became more gradual beyond 15.00 mm in the virtual environment. This discrepancy may be due to HDRI being influenced by nearby light sources such as fluorescent lamps. Light intensity decreases with the square of the distance, so in the real environment, small distance differences can result in significant variations in light intensity. However, HDRI applies fixed lighting regardless of distance, leading to this discrepancy.

In virtual environments with added directional light objects, the simulated grass color characteristics were similar to those in HDRI-only environments, showing an increase up to 15.00 [mm], followed by a more gradual change. However, in the real environment, the OGCD continued to increase up to 17.00 [mm] in Env 4 and up to 20.00 [mm] in Env 5. These results suggest that environments with sunlight may lead to lower accuracy. Due to the limitations of the real-time rendering game engine, accurately simulating sunlight with the HDRI was challenging, so the directional light object was used as a substitute. Environments with sunlight might yield more accurate results if sunlight could be directly calculated with the HDRI, as in 3DCG offline-rendering software. Additionally, adjusting parameters sensitive to strong light, such as metallic and smoothness, and setting BRDF (Bidirectional Reflectance Distribution Function) \cite{guarnera2016brdf} could also potentially improve this issue.

\vspace{-10pt}

\subsection{Comparative Experiment}
\vspace{-5pt}
\subsubsection{Experimental Setup}
We conduct a comparative experiment between our method and the conventional method proposed by Mizuno et al. \cite{mizuno20233d}, which uses a 3DCG offline-rendering software and a genetic algorithm. The experiment was conducted in the same classroom as described in Subsection \ref{multiview}, targeting the grass color characteristics from the position \((h, d, \theta) = (170, 2, 0)\). The grass color characteristics in the real environment and our virtual environment were based on the results from Subsection \ref{multiview}. 

Mizuno et al.'s simulation system was implemented using a 3DCG offline-rendering software (Houdini), a GPU renderer (Redshift), and a genetic algorithm (PyGAD) as described in their previous research paper. The specifications of the computer used for Mizuno et al.'s simulation were as follows: CPU: Core i7-11700, GPU: GeForce RTX 4070, RAM: 32 GB.

\vspace{-15pt}

\subsubsection{Result and Discussion}
\begin{figure}[h]
  \centering
   \vspace{-5pt}
    \includegraphics[width=0.7\linewidth]{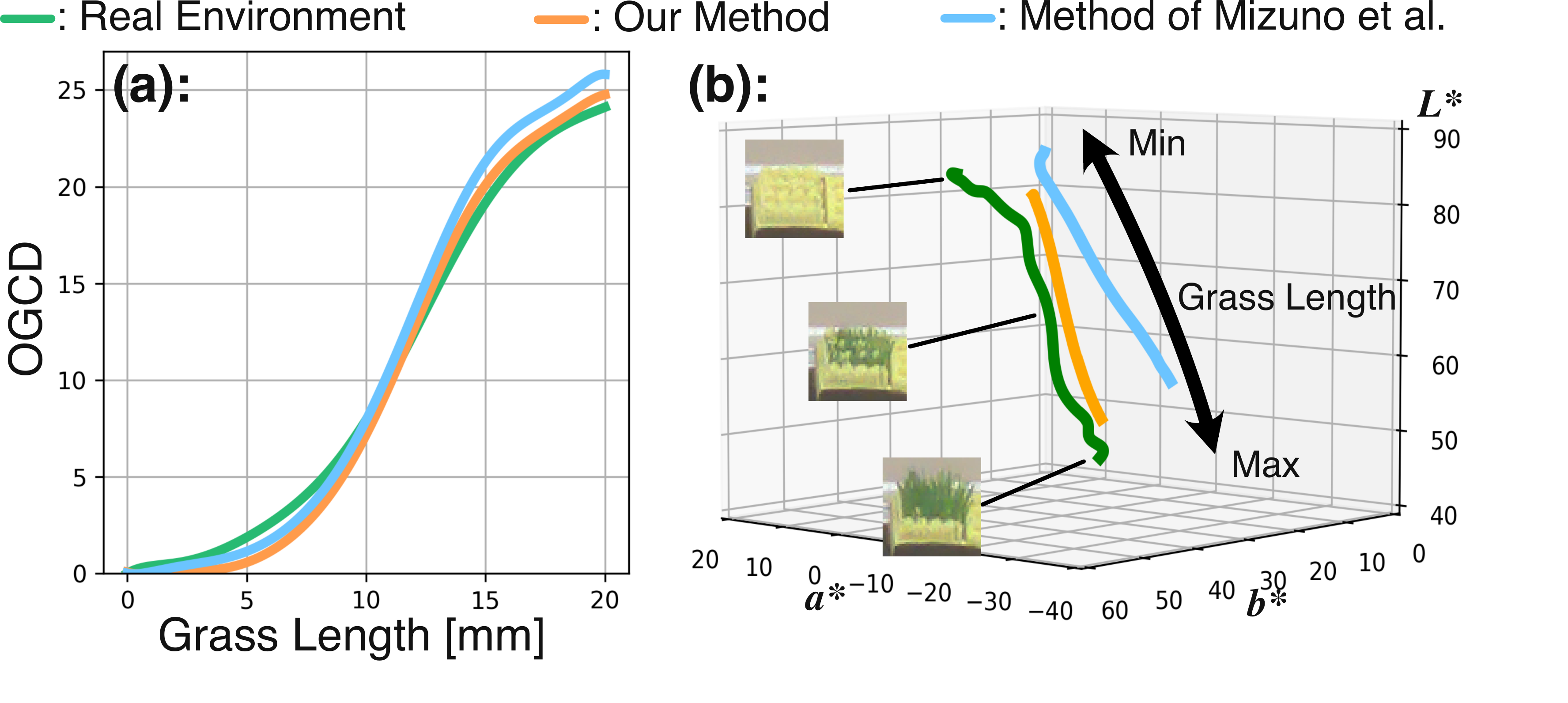}
  \vspace{-15pt}
  \caption{Result of Comparative Experiment} 
  \label{mizunoresult}
  \vspace{-20pt}
\end{figure}

Figure \ref{mizunoresult} presents the result of the comparative experiment. The data obtained from the real environment and our virtual environment are represented in green and yellow, respectively, while the results from Mizuno et al.'s method are shown in blue.
   
In the plot of Figure \ref{mizunoresult}(a), both our method and Mizuno et al.'s method yield results that closely resemble the grass color characteristics of the real environment. The Fréchet distance between the real environment's grass color characteristics and those obtained by Mizuno et al.'s method was 1.71, whereas our method achieved a smaller Fréchet distance of 1.16. Additionally, the plot in Figure \ref{mizunoresult}(b) shows that the CIELAB values of the grass colors from our method and Mizuno et al.'s method were close to each other.

However, our method can obtain the grass color characteristics in several seconds as the grass length changes, whereas Mizuno et al.'s method took about 40 hours of computation time. This significant time difference is due to Mizuno et al.'s use of offline rendering to simulate accurate light calculations and the genetic algorithm to adjust the albedo colors of the fixed-length and adjustable-length grass to match the color of the grass in the real environment as closely as possible. Therefore, the experimental results suggest that even the simulation using real-time rendering of simplified light calculations and setting albedo colors based on the color measurement tool can potentially yield comparable accuracy and faster outcomes compared to simulations with more precise calculations.

Although our method showed better accuracy in this experiment, the accuracy of the offline rendering and the genetic algorithm's results can vary depending on the parameters used and the system on which it is run. Hence, the accuracy results should be considered as reference data.

\vspace{-15pt}

\section{Potential Application}
\vspace{-10pt}
In this section, we explain the potential applications of the grass color characteristic interactive simulation. 
\vspace{-10pt}

\subsection{Virtual Workspace for  Color-Aware  Application Development}

\begin{figure}[h]
  \centering
  \vspace{-20pt}
    \includegraphics[width=\linewidth]{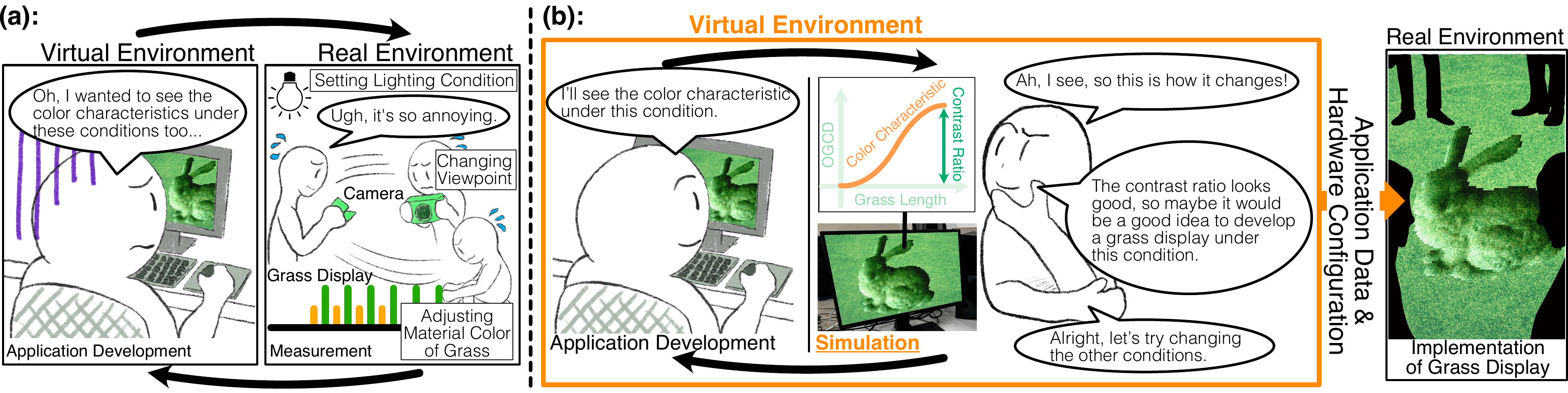}
    \vspace{-20pt}
    \caption{Virtual Application Development Workspace Without (a) and With (b) Grass Display Color Characteristic Simulation (The image was created using the Stanford Bunny 3D model \cite{stanfordbunny} convey the concept of the virtual workspace.)}
  \label{application}
  \vspace{-15pt}
\end{figure}
\vspace{0pt}

 Virtual workspaces, commonly used for developing motion-based media through trial and error \cite{yasu2022magneshape,iwamoto2022playful}, can also support the development of grass display applications. In addition to these basic capabilities, our interactive simulation can support a virtual workspace where developers can obtain the grass color characteristics during the application design process.

To create applications that are accessible and easy to understand for a wide range of users, it is important to understand the color performance of the display device. For example, the World Wide Web Consortium (W3C) provides the Web Content Accessibility Guidelines, which recommend sufficient contrast ratios to improve readability and usability \cite{wcag22}. Since conventional emissive displays such as LCDs maintain stable color performance under changes in lighting and viewpoint, developers can usually focus on designing applications without paying constant attention to such environmental factors.

In contrast, material-based displays such as the grass display are more sensitive to external conditions. Their perceived color characteristics can vary depending on the lighting environment and viewing direction. As shown in Figure 11(a), in a usual virtual workspace, developers can control the physical movement of the grass, but to observe changes in color characteristics under different environmental conditions, they must still perform manual measurements whenever lighting or viewpoint changes. As a result, it can potentially become difficult to develop color-aware applications for the grass display.

As demonstrated in Figure 11(b), a key advantage of our proposed method is the ability to integrate real-world simulations of grass color characteristics into the virtual workspace.  This allows developers to iteratively confirm the display’s color performance based on real-world conditions while developing their applications, including factors such as contrast ratio within the OGCD range. Therefore, our simulation can potentially support effective color-aware grass application development and help developers make more informed decisions before deployment in real environments. 

\vspace{-15pt}

\subsection{Calibration of Grass Display for 8-bit Color Control in Virtual Environment}
\vspace{-8pt}

 On conventional display devices, the design of interactive content is based on stable color control, particularly through the use of 8-bit color levels, which enables clear and accurate rendering of interactive experiences. Although such 8-bit linear control of color is also possible on the grass display \cite{tanaka2024programmablegrass}, the color characteristics of grass vary depending on environmental factors. As a result, achieving accurate color calibration typically requires labor-intensive manual measurements tailored to each deployment context. To address this,  the simulated grass color characteristics can be used not only to verify the performance of the grass display but also to create calibration data for controlling the grass colors at the 8-bit level. By assuming a linear relationship between the 8-bit level and the OGCD based on the grass color characteristics, the relationship between the grass length and the 8-bit level is determined. This allows for the control of the grass colors at the 8-bit level by reflecting this relationship in the grass display. 

 \begin{figure}[h]
  \centering
  \vspace{-15pt}
    \includegraphics[width=0.8\linewidth]{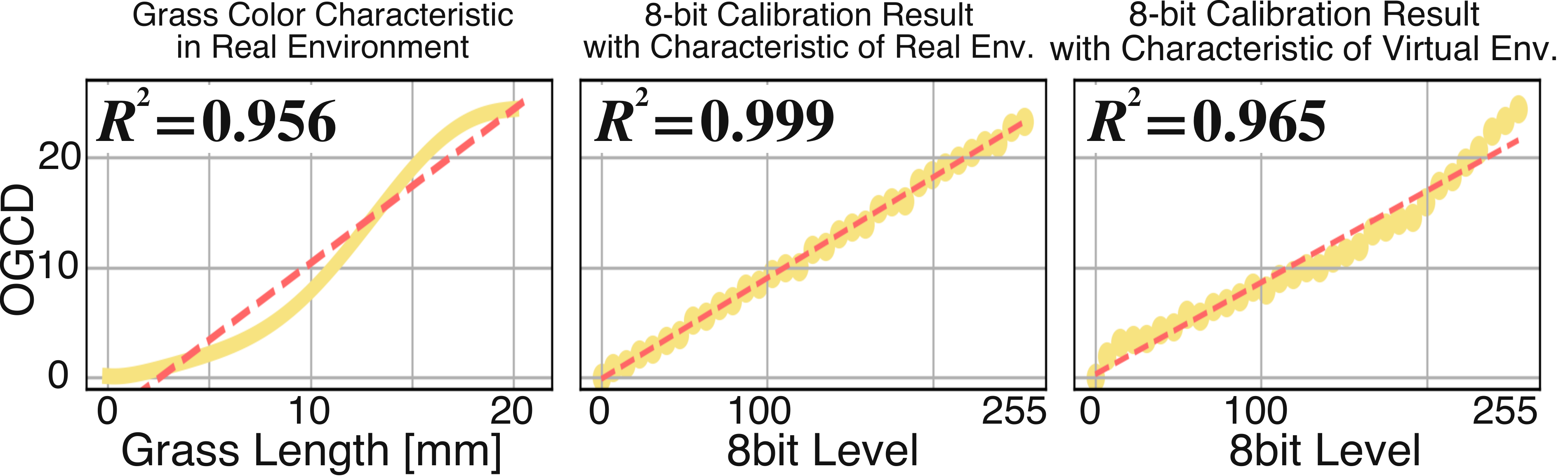}
    \vspace{-10pt}
  \caption{Results of Simple 8-bit Calibration Experiments} 
  \label{calibresult}
  \vspace{-20pt}
\end{figure}

As simplified experiments, the 8-bit calibration of the grass pixel was performed at the position $(h, d, \theta) = (170, 2, 0)$ in the same classroom as Subsection \ref{multiview}. In Figure \ref{calibresult}, the left shows the grass color characteristic in the real environment, the center shows the calibration result based on the real environment characteristic, and the right shows the calibration result from the virtual environment. Linear regression was performed on each plot, with the estimated linear model indicated by a red dotted line and the coefficient of determination $R^2$, which indicates linearity, also calculated.

When the 8-bit calibration was performed using the grass color characteristic from the real environment, the coefficient of determination $R^2$ improved from 0.956 to 0.999. Although the effect was smaller compared to the real environment results, the calibration using the grass color characteristic from the virtual environment also allowed the grass pixel to be controlled linearly, with $R^2$ improving from 0.956 to 0.965. This result demonstrated the potential of the 8-bit grass color calibration in the virtual environment. However, it should be noted that this experiment was conducted under limited conditions, and further validation of the 8-bit calibration effect under various conditions is necessary.

\section{Limitation and Future Work}
\vspace{-10pt}
\subsection{Simulation Experiments}
\vspace{-5pt}
Our method demonstrated the potential to simulate grass color characteristics based on real environments through multiple experiments; however, several limitations remain. It is unclear whether the same level of accuracy can be achieved when different albedo colors are used compared to the grass pixels employed in this study. The color of grass material varies, including various greens and yellows, and recently, artificial grass in pink or blue has also been seen to match interior designs. Therefore, to address this issue, we plan to conduct simulation experiments based on grass pixels of diverse colors.

In this paper, a classroom with stable lighting conditions was chosen as the experimental environment for multi-viewpoint experiments and the comparative experiment with the conventional method. However, the results could differ in other environments. Additionally, while the multiple environment experiments in Sub-section \ref{submultienv} were conducted in relatively bright standard conditions, the accuracy might vary under extreme lighting conditions such as sunset.  We will conduct simulation experiments in various environments, including challenging lighting conditions, to clarify the range of applications for our method.

Multi-viewpoint experiments in outdoor environments are important, however, it is challenging to conduct stable experiments due to changing weather conditions. To address this, indoor environments that are bright and equipped with high-CRI light sources, allowing for HDRI shooting with an omnidirectional camera while maintaining a sufficient distance from the light source, may be feasible. 

In this study, we focused on a small-scale environment with a single grass pixel to make it easier to conduct a multi-faceted evaluation of the effectiveness of the interactive grass display color change simulation, including comparisons across multiple viewpoints, environments, and with previous research. In future work, to further clarify the generalizability of the proposed method, we plan to conduct larger-scale experiments with more grass pixels.

\vspace{-10pt}

\subsection{Design and Implementation}

\vspace{-5pt}

 In this study, we used fixed values for the metallic and smoothness parameters of the 3D grass pixel.  To improve simulation accuracy, not only the albedo but also the metallic and smoothness parameters should be adjusted, and considering the BRDF might also be necessary.  Differentiable rendering, which allows estimation of physical parameters from multiple images, may be useful for efficiently acquiring such properties.  Additionally, the grass length of the virtual grass pixel was adjusted by moving the relative position of the green grass. However, considering physical simulations such as collision detection and gravity may improve the accuracy of simulating the grass color characteristics.

In this paper, since it was difficult to reproduce direct sunlight using HDRI in the interactive virtual environment, we substituted it with a virtual directional light object. However, technologies that can replicate sunlight using only HDRI in real-time rendering are currently being developed \cite{truehdri}. We plan to use these technologies to improve the accuracy of outdoor simulations.

The Unity HDRP was used with mostly default values, except for HDRI and directional light objects. However, it is necessary to investigate whether the use of features such as subsurface scattering, which allows for more photorealistic representation, can achieve better results. Moreover, we chose the Unity HDRP to accurately calculate colors based on physical properties; however, this system requires a high-performance computer. In the future, to clarify the versatility of the simulation, we will conduct experiments to verify how the simulation results differ when using a game engine with the standard pipeline. 

In this paper, we focused on the color change simulation of the grass display. However, we believe there is potential for our proposed method to be applied to simulate color changes in media using other everyday materials by adjusting the material color and shape to match the specific requirements of these media. Based on the insights gained from this study, we will explore color characteristic simulations for media using other everyday materials.

\vspace{-12pt}

\section{Conclusion}
\vspace{-7pt}
We explored an interactive simulation method for grass display color characteristics based on real conditions. Our focus was on balancing accuracy with ease of preparation to achieve an effective interactive simulation of the grass color characteristic. The main contributions of this paper are summarized as follows:

 \begin{itemize}
  \item We designed and implemented an interactive simulation of the color characteristics in the grass display based on real-world conditions (Figure \ref{colorprocess}).
  
  \item  Our experiments demonstrated the potential of the proposed interactive method to reliably simulate grass color characteristics across multiple viewpoints and environments  (Figure \ref{charactresult}, \ref{multienv}).
  
  \item Our simulation method showed the potential to simulate with comparable accuracy to the previous study while achieving faster speeds (Figure \ref{mizunoresult}).
\end{itemize}

\noindent In addition, we had simple discussions on the potential applications of the proposed simulation for implementing the grass display in real-world environments. 

\vspace{5pt}

In future work, we will improve the simulation method to maintain accuracy across a wider range of viewpoints. To achieve this, we plan to develop a grass display simulation that considers two primary limitations: one is the gradual decrease in accuracy as the viewpoint moves away from the reference position used for color correction; the other is  the drop in accuracy at viewpoints where the slits of the display become more visible, which may result from differences in the appearance of the grass caused by physical behavior between the real and virtual environments. In addition to these improvements, we will  also  conduct simulation experiments that examine various grass material colors and challenging lighting conditions to further verify the generalizability of the grass color characteristic simulation.

\section*{Acknowledgment}
 This work was supported by JSPS KAKENHI Grant Number JP24KJ0463. 

\bibliographystyle{unsrt}

\bibliography{camera-ready4}
\end{document}